\documentclass[twocolumn,10pt]{asme2ej}

\usepackage{graphicx} 

\usepackage{array}
\usepackage{amsfonts}

\usepackage{placeins}
\usepackage{stfloats}

\usepackage{background}
\backgroundsetup{color=gray}

\title{Towards Reusable Surrogate Models: Graph-Based Transfer Learning on Trusses}

\author{Eamon Whalen\thanks{Address all correspondence to this author.}
    \affiliation{
	Graduate Research Assistant\\
	Computational Science and Engineering\\
	Massachusetts Institute of Technology\\
	Cambridge, Massachusetts 02139\\
    Email: ewhalen@mit.edu
    }	
}

\author{Caitlin Mueller
    \affiliation{Associate Professor\\
	Civil and Environmental Engineering\\
	Massachusetts Institute of Technology\\
	Cambridge, Massachusetts 02139\\
    Email: caitlinm@mit.edu
    }
}

\SetBgContents{\ifnum\value{page}=1 Accepted manuscript, ASME Journal of Mechanical Design. doi:10.1115/1.4052298 \else \fi}
\SetBgScale{1}
\SetBgAngle{0}
\SetBgPosition{current page.north west}
\SetBgHshift{10cm}
\SetBgVshift{3cm}

\begin{document}

\maketitle    

\begin{abstract}
{\it Surrogate models have several uses in engineering design, including speeding up design optimization, noise reduction, test measurement interpolation, gradient estimation, portability, and protection of intellectual property. Traditionally, surrogate models require that all training data conform to the same parametrization (e.g. design variables), limiting design freedom and prohibiting the reuse of historical data. In response, this paper proposes Graph-based Surrogate Models (GSMs) for trusses. The GSM can accurately predict displacement fields from static loads given the structure’s geometry as input, enabling training across multiple parametrizations. GSMs build upon recent advancements in geometric deep learning which have led to the ability to learn on undirected graphs: a natural representation for trusses. To further promote flexible surrogate models, the paper explores transfer learning within the context of engineering design, and demonstrates positive knowledge transfer across data sets of different topologies, complexities, loads and applications, resulting in more flexible and data-efficient surrogate models for trusses.
}
\end{abstract}

\begin{nomenclature}
\entry{DM}{Design Model}
\entry{GNN}{Graph Neural Network}
\entry{GSM}{Graph-based Surrogate Model}
\entry{MAE}{Mean Absolute Error}
\end{nomenclature}

\section{Introduction}
Surrogate models, also known as metamodels, response surfaces, reduced order models, approximation models, or emulators, are used extensively in engineering to approximate complex systems. In a typical workflow, training data is produced by running a design of experiment (DOE) of physics-based simulations, after which a surrogate model is trained in a supervised manner to predict one or more of the simulated quantities. The trained surrogate model might then be used to speed up design optimization, estimate gradients (if the surrogate is differentiable), or share the system (e.g. to a lightweight/web platform or collaborator) easily and without divulging intellectual property. Generally, these surrogate modeling methods require that each design be represented as a fixed-length vector of design parameters (e.g. design variables). This requirement restricts the surrogate model to a single design space, requiring the user to train a new surrogate model every time the parametrization changes.

Ideally, surrogate models would operate on more organic representations of geometry, enabling learning across design data from multiple sources. Many design processes are incremental in nature. The result is often several small, disjoint design studies which differ slightly in geometry, topology, or loading conditions. A more flexible surrogate model could be trained across design iterations, perhaps supplemented with historical designs from previous projects, and could be continuously updated as new data becomes available. The ability to learn across related projects would not only save computational resources but might also yield powerful insights that could not have been inferred from a single design space. Such models would also grant engineers greater design freedom since design changes would not be restricted to the parametrization used to generate training data.

One challenge in developing such a model is choosing a geometry representation. An ideal representation would accommodate arbitrary changes to the geometry or topology, and encode loads and supports to enable learning across load cases. A second challenge is quantifying the extent to which such a surrogate generalizes to new designs. Unlike with traditional parametrization, the notions of interpolation and extrapolation are not well-defined for representations that span the set of all possible shapes. How might one determine which inputs are “safe” and which are not likely to yield quality predictions?

This work explores the use of graph neural networks as surrogate models for trusses. The proposed Graph-based Surrogate Model (GSM) learns to predict a displacement field given only the geometry, supports, and loads as inputs. It is shown that the GSM can be trained on data from multiple design models simultaneously, often outperforming GSMs trained on a single source. Transfer learning is then explored as an effective method to repurpose previously trained GSMs to new tasks. Both the generalizability and data efficiency of the GSM are improved with transfer learning, with positive transfer being observed across varying topologies, loads, complexities, and even different applications.

The key contributions of this work are as follows:
\begin{enumerate}
  \item Graph-based Surrogate Models (GSMs), which operate directly on the geometry and do not require parametric design features, are proposed for the modeling of trusses
  \item Transfer learning is shown to improve the GSMs data efficiency and generalizability, leveraging historical data to reduce the required number of simulations by one or two orders of magnitude
  \item Various source/target pairs that arise naturally in a design context, including design data of varying topologies, loads, complexities and applications, are used to demonstrate the utility of transfer learned GSMs in a real world setting
\end{enumerate}

The remainder of this paper is organized as follows: section \ref{relatedWork} reviews related work, section \ref{meth} introduces the methodology of the GSM and a few naive alternatives used for comparison, section \ref{exp1} outlines data generation methods and presents experimental results, section \ref{exp2} introduces transfer learning and presents further results, and section \ref{conc} contains conclusions and ideas for future work.

The following terminology is used throughout the paper: Let \emph{design} refer to a specific design concept of a structure (i.e. something that could be built), \emph{design model} (DM) refer to a hand-parametrized design space which can be sampled to generate designs, and \emph{surrogate model} refer to a data-driven predictive model that learns to predict a structure’s engineering performance.

\section{Related work} \label{relatedWork}
Engineering surrogate modeling is a thoroughly explored topic with applications dating back to the 1980s. Conversely, transfer learning and geometric deep learning are relatively young research areas with hundreds of papers published in the last few years alone. The following is a brief review of what are considered to be the most relevant works to this one, but is by no means comprehensive.

\subsection{Surrogate modeling with parametric design features}
Surrogate models have been used in engineering design for several decades (see \cite{wang_review_2007, forrester_engineering_2008, queipo_surrogate-based_2005} for a review). Some of the most common surrogate modeling algorithms include polynomial regression \cite{sacks_design_1989}, Kriging (also known as Gaussian processes) \cite{cressie_spatial_1988}, radial basis functions \cite{dyn_numerical_1986}, random forest \cite{tin_kam_ho_random_1995} and neural networks \cite{papadrakakis_structural_1998}. \cite{tseranidis_data-driven_2016} compared several of these algorithms for civil engineering problems. Dimensionality reduction techniques have been used to derive more suitable parametrizations \cite{brown_design_2019, danhaive_design_2021} and quantities of interest \cite{xu_multi-level_2020}. All of the aforementioned methods require that a design be represented as a fixed-length vector of parametric design features, restricting the feasible designs to some pre-determined space. This work proposes a surrogate model that operates on the geometry directly and is thus not limited to a particular parametrization.

\subsection{Surrogate modeling without parametric design features}
Recently, a few surrogate models have been proposed that do not rely on handcrafted design parameters. \cite{xuereb_conti_flexible_2018} proposed using "knowledge-based" characteristics, which are independent of design variables, as features. While this may enable the combination of training data from multiple design spaces, it still relies heavily on the user to craft useful characteristics. Other approaches, have sought to learn on the geometry itself. The pursuit of deep learning methods for shape data has led to the ability to learn on several geometry representations, including shape descriptors, images, voxels, polycubes, signed distance functions, point clouds, and graphs (see \cite{liu_deep_2020, ahmed_survey_2019} for a review). Surrogate models have been trained on images \cite{jiang_stressgan_2020, messner_convolutional_2020, yoo_integrating_2021, madani_bridging_2019, garland_pragmatic_2021, guo_convolutional_2016}, voxels \cite{zhang_featurenet_2018, williams_design_2019} and polycubes \cite{umetani_exploring_2017, baque_geodesic_2018}.
Images and voxels suffer from resolution problems and data loss due to rasterization. Polycubes solve this problem by mapping the geometry to a regular grid but are limited to fixed-topology data sets. 

The advent of geometric deep learning techniques has enabled learning on non-Euclidian domains which are generally more natural representations of geometry. \cite{cunningham_investigation_2019} trained a surrogate model to predict lift and drag coefficients from 3D point clouds. While potentially useful for solid bodies, point clouds are not an adequate representation of trusses because they lack topological information. Other works have represented designs as graphs. \cite{baque_geodesic_2018} used a graph-based convolutional model to learn fluid dynamics on meshed surfaces, \cite{danhaive_structural_2020} used a similar approach to learn the structural behavior of a thin shell, and \cite{vlassis_geometric_2020} learned material properties from graph-based microstructures. The closest existing work to this one is probably \cite{chang_learning_2020}, in which graph representations of trusses were used to optimize cross section sizes for structural loads. The structures in \cite{chang_learning_2020} had constant loads and geometry (apart from the cross sections), whereas this study explores the flexibility of graph-based networks to generalize across various geometries, topologies and loads.

Other notable engineering applications of graph-based learning include feature recognition on 3D CAD \cite{cao_graph_2020}, shape correspondence for additive manufacturing \cite{huang_geometric_2020}, and generation of design decision sequences \cite{raina_learning_2019}; however, these do not directly address surrogate modeling.

\subsection{Geometric deep learning: learning on graphs}
Graph-based learning, both for shape analysis as well as other tasks, has recently received a lot of attention. \cite{bronstein_geometric_2017} introduced the term \emph{geometric deep learning} to mean learning from non-Euclidian data structures such as graphs and point clouds. See \cite{wu_comprehensive_2020, zhou_graph_2019} for a general survey on graph neural networks (GNNs). MoNet \cite{masci_geodesic_2018} was the first framework to apply a GNN to meshed surfaces by leveraging convolutions over local geodesic patches. ACNN \cite{boscaini_learning_2016} defined similar patches based on anisotropic heat kernels, while GCNN \cite{monti_geometric_2016} generalized these patches to user-defined pseudo coordinates. FeaStNet \cite{verma_feastnet_2018} introduced an attention mechanism to perform "feature steering" which acts as dynamic filtering over neighbors. Other notable extensions of GNNs to shapes include MeshCNN \cite{hanocka_meshcnn_2019} which introduced learnable edge pooling and StructureNet \cite{mo_structurenet_2019} which introduced a graph-based encoder for hierarchical part representations. The aforementioned frameworks were applied to geometry processing tasks including shape correspondence, classification, and segmentation, whereas this work focuses on structural surrogate modeling.  

\subsection{Transfer learning: Recycling data}
Transfer learning, where predictive models previously trained on source data are re-trained on target data from a different domain, task, or distribution, is a widely applied concept in machine learning \cite{pan_survey_2010}. Deep learning models in particular often benefit from transfer learning due to their data-intensive nature \cite{tan_survey_2018}. \cite{lee_transfer_nodate} addressed some of the particular challenges of transfer learning in graph neural networks. A few works have explored transfer learning in the context of engineering design. \cite{yoo_integrating_2021} trained a convolutional autoencoder on 2D wheel designs before retraining the encoder as a surrogate model, reducing the required number of simulations. \cite{madani_bridging_2019} first trained a model to predict the original parametric design features of an artery before retraining it to predict the location of maximum stress. \cite{li_efficient_nodate} used a clustering algorithm to identify which designs would make for useful source data when applying transfer learning to microprocessor performance prediction. \cite{baque_geodesic_2018} trained a surrogate model to predict the drag coefficient of 2,000 primitive shapes before tuning the model on 54 car designs. This paper differs from previous works in that it seeks to systematically quantify the effects of transfer learning on data efficiency and generalizability across several common source/target pairs in structural design.

\section{Methodology: Surrogate modeling with graphs} \label{meth}
The following section presents a new graph-based surrogate model (GSM) for predicting the displacement of trusses under static load.

\subsection{Data representation: Trusses as graphs}
This paper proposes a graph-based representation of trusses, where a set of vertices \(V = \{v_1, ..., v_n\}\) represent the joints and a set of edges  \(E \in V \times V\) represent the bars. The set of vertices that share an edge with \(v_i\) is referred to as its neighborhood, and is understood to include \(v_i\) itself. Each vertex \(v_i\) is assigned a feature vector \(x_i\) of length \(r\) (\(x_i \in \mathbb{R}^r\)). The geometry of the truss is encoded by using the joints’ spatial coordinates \(c_i \in \mathbb{R}^2\) as vertex features. Additional binary features indicate the presence of a support \(s_i \in \{0,1\}^2\) or load \(l_i \in \{0,1\}^2\) for each degree of freedom. The geometry, supports and loads are thus encoded by the graph \(G_0 = (V_0,E)\). The deformed structure is represented by a topologically identical graph \(G_H = (V_H,E)\), where now the vertex features encode the displacements \(d_i \in \mathbb{R}^2\) of each joint under static load. The proposed graph representation has three main advantages: 
\begin{enumerate}
  \item it encodes the exact spatial coordinates of the geometry
  \item it facilitates arbitrary topologies
  \item it does not rely on handcrafted design parameters
\end{enumerate}
In contrast with Euclidian representations like images, 1. implies that there is no information loss when converting the geometry to or from the deep learning representation. 2. and 3. enable learning across multiple design spaces.

\begin{figure} 
\centerline{\includegraphics[width=3.25in]{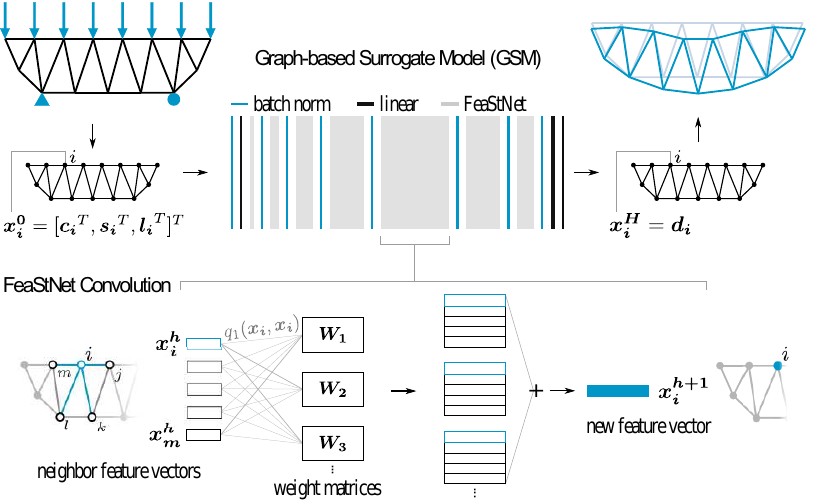}}
\caption{The graph-based surrogate model (GSM) learns to predict nodal displacements given only geometry, supports and loads as inputs. Structures are represented as undirected graphs, where each vertex is assigned a feature vector consisting of a joint’s spatial coordinates and binary variables indicating the presence of supports or loads. Graph convolutional layers utilize the FeaStNet operator \cite{verma_feastnet_2018}.}
\label{architectureFig}
\end{figure}

\subsection{Convolutions on graphs}
The GSM's primary mechanism is a graph-based convolutional layer. The FeaStNet \cite{verma_feastnet_2018} convolution was selected because it extends to arbitrary graph topologies, does not require the selection and pre-computation of pseudo coordinates, and can be made transformation invariant in feature space. The latter implies that raw spatial coordinates can be used directly as input features without having to learn spatial invariance or transform all designs to a common pose. Geometric deep learning is an active field; it is likely that other graph-based learning methods are also suitable for this context and should be considered as future research.

\subsection{The graph-based surrogate model (GSM)}
The proposed surrogate model learns to predict joint displacements given the geometry, supports and loads as inputs. It does so by learning a map from an input graph \(G_0 = (V_0,E)\) to a topologically identical output graph \(G_H = (V_H,E)\). The surrogate model is implemented as a graph-based convolutional neural network built from a single sequence of \(H\) linear and FeaStNet convolutional layers (Fig.~\ref{architectureFig}). All layers except the final one are followed by a rectified linear (ReLu) activation function. It is observed that batch normalization applied to the input and after each convolutional operation significantly improves prediction accuracy. The network architecture, layer dimensions, and number of attention heads per FeaStNet layer dictate the total number of learnable parameters.

\subsection{A naive alternative: The pointwise surrogate}
A second, simpler type of surrogate model was used to compare against the proposed graph-based method. This \emph{pointwise surrogate} consists of several simple regression models, which each take the spatial coordinates of the structure's joints (flattened into a vector) as inputs and predict a single scalar quantity. For a 2D truss with 15 nodes, this corresponds to training 30 regression models (for the \emph{x} and \emph{y} displacement of each node). The random forest algorithm was selected for this study, but any regression technique (e.g. Kriging, polynomials, radial basis functions) could be used. Note that the pointwise surrogate relies on a fixed ordering of joints and thus cannot be extended to multi-topology data sets. Also, note that in the case where all designs are identically loaded, there is no benefit to including support or load information in the input, since the designs are represented by a single vector. The pointwise surrogate was implemented using the scikit-learn \cite{pedregosa_scikit-learn_2011} random forest class using default settings.

\subsection{A baseline: Predicting the mean}
As an additional reference point, consider an even simpler predictive model that simply predicts the mean displacement across each joint in the training set. Throughout the paper, the performance of this naive model is referred to as the \emph{baseline}. Models that fail to beat the baseline effectively have no predictive value.

\section{Characterizing the GSM} \label{exp1}
The following section presents a series of trials designed to characterize the prediction accuracy and generalizability of the proposed graph-based surrogate model.

\subsection{Data generation and filtering} \label{dataGen}
\begin{figure} 
\centerline{\includegraphics[width=3.25in]{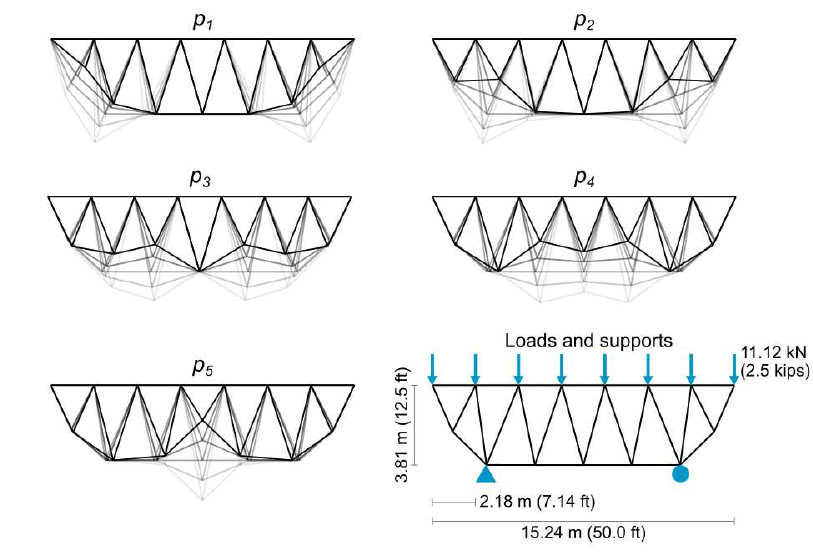}}
\caption{A parametric design model of a truss. Data sets are created by perturbing design variables \(p_1\)-\(p_5\). Each design is loaded with a uniformly distributed vertical load across the top and simply supported on the bottom. This particular design model is referred to as DM7.}
\label{desSpaceFig}
\end{figure}

Surrogate modeling is most advantageous for computationally-intensive simulations; however, this work focuses on relatively simple designs because they more effectively depict the specific design scenarios used to evaluate the GSM (more on this in section \ref{exp2}). A set of truss designs was generated as follows. First, a parametric design model of a simple two-dimensional truss was built using a combination of commercial \cite{rutten_grasshopper_nodate} and open source \cite{huang_pyconmech_2020} software. The truss is made of steel (\(E=\) 30.5 Msi) and consists of beams with constant cross section (\(A=\) 0.29 \(m^2\), \(I=\) 2.3\(\times10^{-3}\) \(m^4\)). A vertical static load of 11.1 kN is applied to all joints on the top of the truss, and simple supports are applied to two of the bottom joints (Fig.~\ref{desSpaceFig}). The truss was parametrized using five handcrafted design variables \(p_1\)-\(p_5\), each perturbing the truss geometry in a particular way. Next, the design model was sampled 1,000 times using a Latin Hypercube and the resulting designs were simulated with bar elements using linear elastic Finite Element Analysis (FEA). Finally, the 10\% of trusses with the largest maximum displacement (i.e. the worst-performing designs) were discarded. For the remainder of the paper, this design model will be referred to a design model 7 (DM7).

\subsection{Training and tuning} \label{training}
A GSM was trained to predict joint displacements given a truss design as input. The truss designs were randomly partitioned such that 68\% were used for training, 12\% were used for validation, and 20\% were reserved for testing. The GSM was implemented with Pytorch Geometric \cite{fey_fast_2019} and trained for 100 epochs on a Tesla K80 GPU using the ADAM optimizer \cite{kingma_adam_2017} and a mean squared error (MSE) loss function. Through a series of grid searches, the optimal architecture was found to be \emph{L16/C32/C64/C128/C256/C512/C256/C128/L64/L2}, where \emph{L} denotes a linear layer, \emph{C} denotes a FeaStNet convolutional layer, and the numbers represent the length of the vertex feature vectors after passing through a given layer. Similarly, the optimal learning rate was found to be 1\(\times10^{-3}\) and the optimal number of FeaStNet heads was found to be 8 (see appendix Table \ref{hyperparamTable} for details). The resulting model has 2.7 million training parameters. Throughout all trials, batch normalization was applied to the input and after each convolutional layer, the ADAM weight decay was set to 1\(\times10^{-3}\), and the batch size was set to 256.

Four data transformation strategies were studied: standardization, log transformation, standardization followed by log transformation, and no transform. It was found that standardization alone yields the lowest testing MSE (appendix Table \ref{transTable}). To study the effects of including support and load information in the feature vectors, the model was trained once using spatial coordinates alone as features and compared to when spatial coordinates are used in addition to binary support or load features. It was found that including both the support and load features in the feature vector improves prediction accuracy, despite the fact that all trusses were loaded identically (appendix Table \ref{featureTable}). This is understandable, since the convolution can be thought of as acting on one vertex at a time.

\subsection{Comparing the GSM to the pointwise surrogate}
\begin{figure} 
\centerline{\includegraphics[width=3.25in]{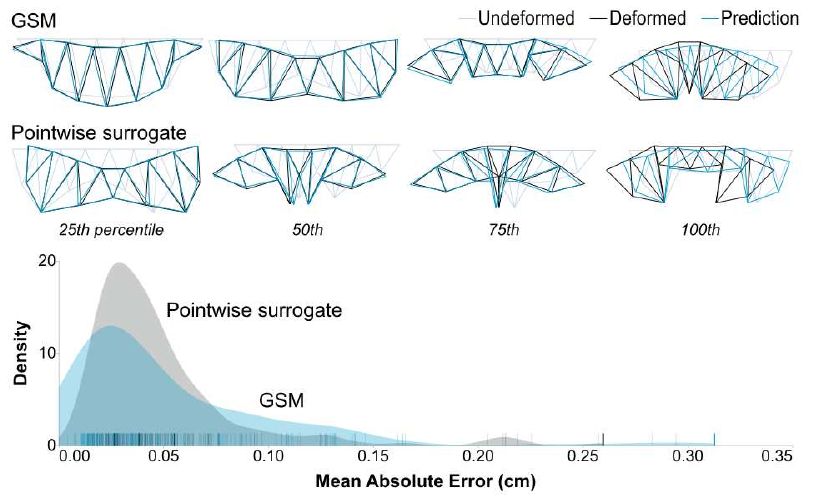}}
\caption{The GSM and pointwise surrogate achieve comparable predictive performance on the test designs. Both error distributions are left-skewed, with 85\% of designs producing a mean average error of less than 0.1 cm on either model.}
\label{gsmPrComp}
\end{figure}

Both the GSM and pointwise surrogate successfully learn to predict a wide range of structural behaviors. Figure \ref{gsmPrComp} shows the distribution of prediction errors for both models evaluated on the test set. The predictive performance of the two models is roughly comparable: the mean absolute error (MAE) over the entire test set is 0.049 cm for the GSM and 0.053 cm for the pointwise surrogate (30\% and 33\% of the baseline respectively). As the average maximum displacement of the trusses is 0.56 cm, these prediction errors are acceptable for most applications. The error distributions for both models are skewed left, implying that the models perform well on most of the designs but poorly on a few. Interestingly, it is observed that many of the designs for which prediction accuracies are low tend to also exhibit poor structural performance (i.e. large displacements). 

\begin{figure} 
\centerline{\includegraphics[width=3.25in]{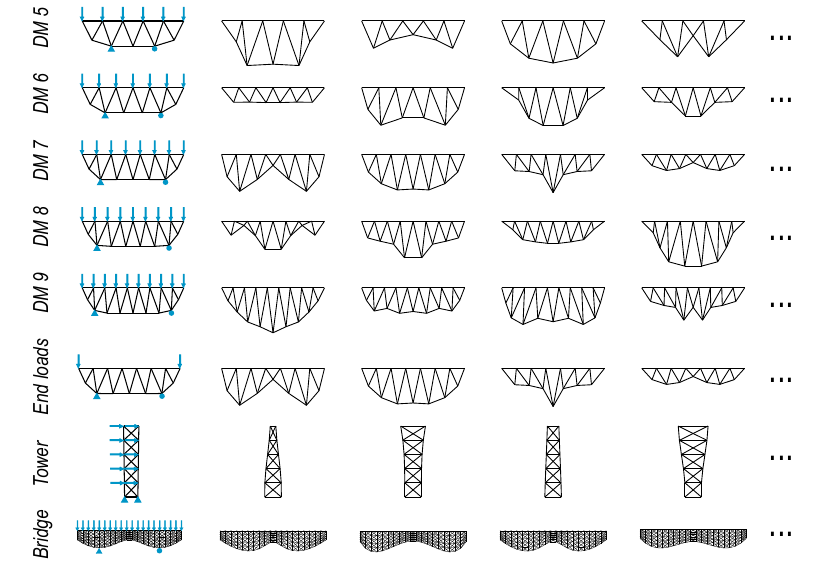}}
\caption{Each row shows a few designs generated from one of the eight design models used in this paper. Loads and supports are omitted on all but the first column for clarity. The design models were selected to test specific scenarios that commonly arise in engineering design.}
\label{multitopoSamples}
\end{figure}

\begin{figure*} 
\centerline{\includegraphics[width=6.5in]{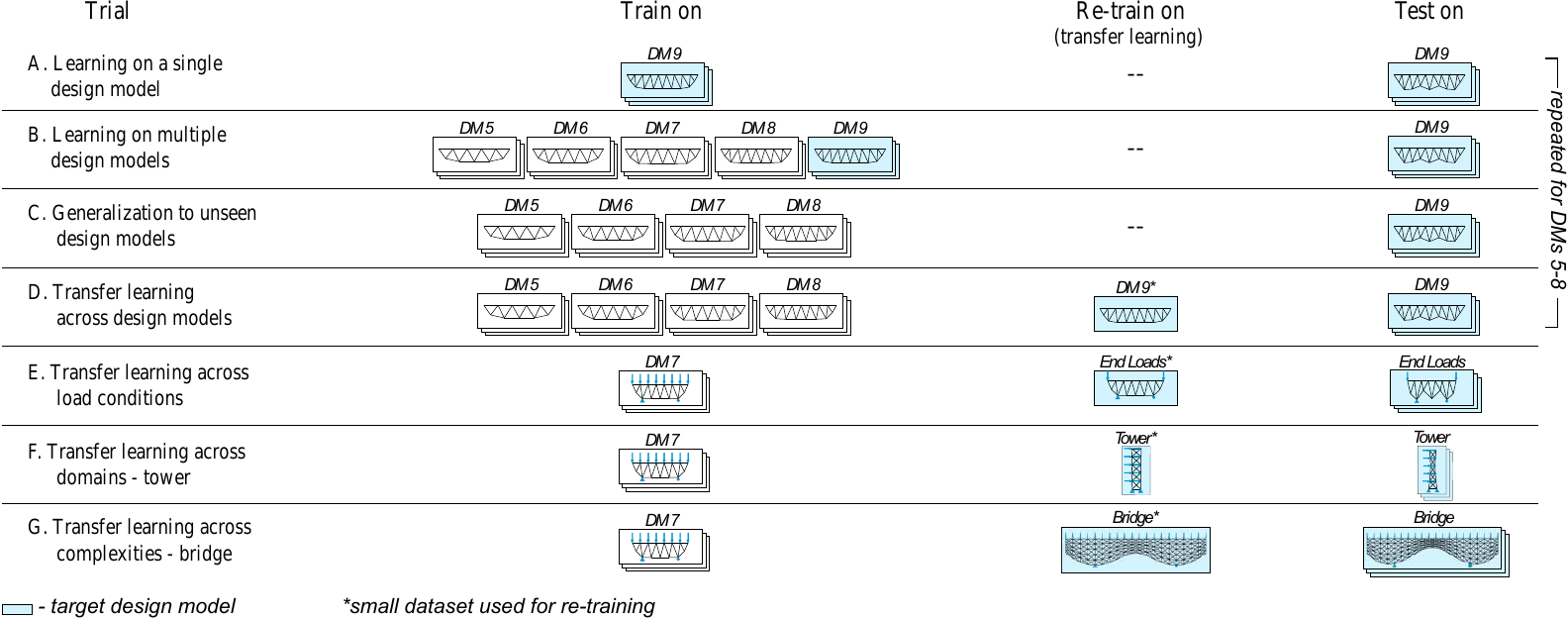}}
\caption{An overview of the trials used to assess the GSM's generalizability across seven specific scenarios. The first three trials involve a single training, while the remainder of the trials leverage transfer learning to repurpose a previously trained GSM for new tasks. Note that trials \emph{\small Trials B-G}) would not be possible with a traditional surrogate model because the data does not conform to the same parametrization.}
\label{trialMap}
\end{figure*}

\begin{figure} 
\centerline{\includegraphics[width=3.25in]{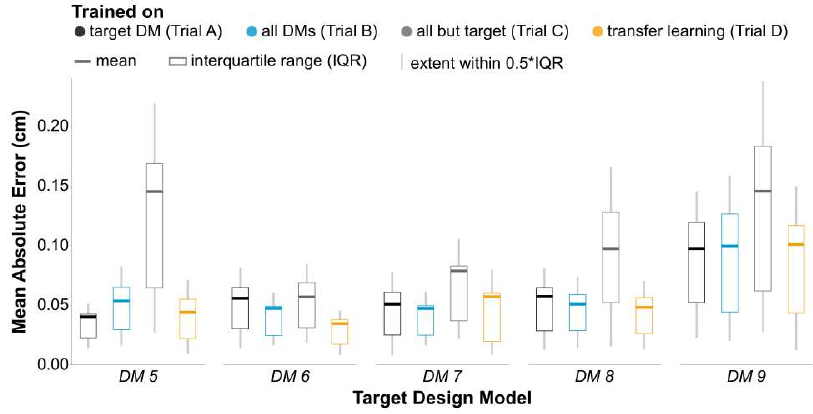}}
\caption{The GSM can learn on data from multiple design models at once (\emph{\small Trial B}), and doing so is sometimes advantageous even for cases when only a single design model is of interest. The GSM does not seem to generalize well to unseen design models (\emph{\small Trial C}); however, transfer learning is an effective remedy (\emph{\small Trial D}) and requires a fraction of the data required to train a GSM from scratch.}
\label{generalizationStudy}
\end{figure}

\begin{figure*} 
\centerline{\includegraphics[width=6.5in]{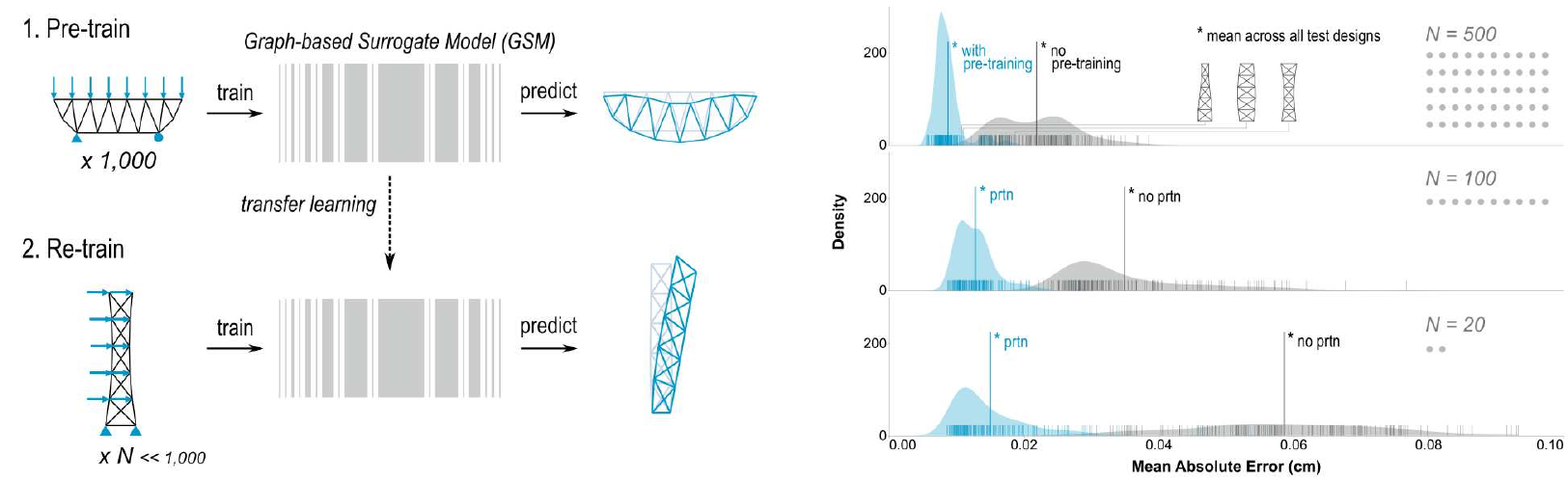}}
\caption{Left: A previously trained Graph-based Surrogate Model (GSM) can be re-trained on a new data set with differing geometry, loads or topology. Right: Pre-training significantly increases the data efficiency of the GSM. In these results from \emph{\small Trial F}, a pre-trained GSM trained on 20 designs (\emph{\small N}=20) outperforms a fresh GSM trained on 500.}
\label{overview}
\end{figure*}

\subsection{Studying generalizability}
Effective surrogate models should generalize well to unseen designs. For surrogate models that rely on bounded, handcrafted design parameters, one might assess generalizability simply by sampling the design space with sufficient density. In contrast, graph representations span the set of all conceivable trusses and thus a bounded design space does not exist.
Developing practical intuition regarding the extent to which graph-based surrogate models generalize to new designs is an open challenge.

Towards this end, a series of data sets and trials were designed to test the generalizability of the GSM under a variety of conditions. The truss design model from section \ref{dataGen} (DM7) was modified to create four new design models. The new design models, named DM5, DM6, DM8, DM9 for the number of bars along the top, have identical outer profiles as DM7 but differing topologies (Fig.~\ref{multitopoSamples}). 

The following trials were designed to test the generalizability of the GSM. The reader is referred to Figure \ref{trialMap} for an overview of the trials used throughout the rest of the paper. Let the term \emph{target} refer to the design model of interest to the user, that is, the design model from which the test set was generated. In \emph{Trial A}, a GSM was trained and tested on designs generated from the target design model. Note that there is no overlap between the training and testing sets. In \emph{Trial B}, training data from all of the design models was combined to train the GSM. The GSM was then tested on designs from the target design model as in \emph{Trial A}. \emph{Trial B} thus quantifies the GSM’s ability to learn on multiple design models simultaneously. Note that this would be impossible with the pointwise surrogate which is limited to fixed-topology data. In \emph{Trial C}, designs originating from the target design model were removed from the training set, thus testing the GSM’s ability to generalize to unseen design models. \emph{Trials A-C} were repeated with each of the five design models (DMs 5-7) as the target, the results of which can be seen in Figure \ref{generalizationStudy}.

In \emph{Trial A}, the GSM archives a MAE of less than 0.1 cm for all design models, confirming the previous conclusion that the GSM effectively approximates single design model data. \emph{Trial B} also produced MAEs less than 0.1 cm across each design model, indicating that the GSM can learn on data from multiple design models simultaneously. Interestingly, for three of the design models (DMs 6-8), the inclusion of data from other design models actually improved predictions on the target. These results indicate that it is sometimes beneficial to add designs to the training data even if they are not from the design model of interest. Note that this did not hold true for DMs 5 or 9 which might be considered the most different from the rest of the design models in that they have the fewest and most bars, respectively. The degree to which including off-target designs in the training data benefits training may therefore depend on how similar those designs are to the target. 

Since the GSM is able to learn on multiple design models simultaneously, one might hope that the model generalizes well to previously unseen design models; however, this was not the case. The MAEs produced in \emph{Trial C} were on average 76\% higher than those in \emph{Trial B}. While the mid-range topologies (DMs 6-8) showed better generalization than the extremes (DMs 5,9), the general trend was that removing all target designs from the training data significantly reduces predictive performance. In other words, the GSM does not seem to generalize well to unseen design models. It is possible that greater topological variation in the training set is required in order to learn such generalization.

\section{Transfer learning: repurposing the GSM} \label{exp2}
In light of the GSM's poor generalization to unseen design models, one might conclude that a separate GSM must be trained for each potential target; luckily this is not the case. This paper proposes transfer learning as a means of repurposing previously trained GSMs to new targets, using a fraction of the training data required to train a GSM from scratch. Transfer learned GSMs thus reduce the number of required simulations and training epochs, and enable learning across design models, load conditions, and even separate applications. This section demonstrates the benefits of applying transfer learning to GSMs through four trials that emulate common design scenarios. Namely, learning across small data sets which vary in topology, loads, application, or complexity.

\begin{figure} 
\centerline{\includegraphics[width=3.25in]{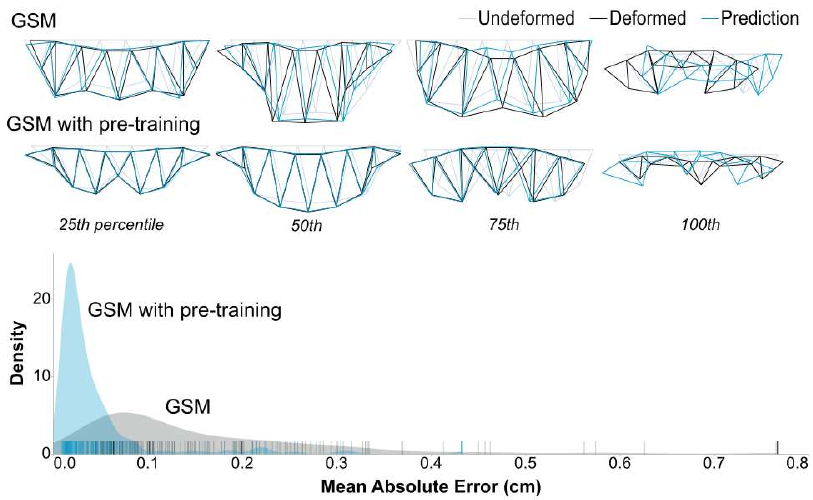}}
\caption{A comparison of prediction error distributions between a pre-trained GSM and a GSM trained from scratch (\emph{\small Trial D}, DM7). Both GSMs were trained on \emph{\small N}=200 target designs. Visualizations of predictions representing each error quartile are also shown. Transfer learning reduces both the mean and standard deviation of prediction errors across the test designs.}
\label{tlVizComp}
\end{figure}

\subsection{Effects on generalizability}
Consider a GSM that has been trained to predict the performance of one or more design models as described in sections \ref{meth} and \ref{exp1}. Let these design models now be referred to as the \emph{source}. The subsequent trials demonstrate the performance of this GSM when re-trained on a small training set from a new target design model (Fig.~\ref{overview}). Multiple strategies exist for applying transfer learning to neural networks. This study employs what is perhaps the most basic: simply retraining all learnable parameters on the target data set for an additional 100 epochs. The further exploration of transfer learning strategies, for example those that freeze parameters or add new ones, for engineering design encouraged as future work. 

In \emph{Trial D}, the GSMs that were previously trained (pre-trained) on all design models but the target (\emph{Trial C}) are re-trained on a small dataset (\emph{N}=200) from the target model. The results can be seen in the final series of Figure \ref{generalizationStudy}. The re-trained GSMs produce significantly better predictions than those in \emph{Trial C}. In fact, the re-trained GSMs on average produce 5.5\% lower errors than those in \emph{Trial B} and use less than a third of the training data. To further analyze the effects of transfer learning on prediction accuracy, the error distribution from the pre-trained GSM in \emph{Trial D} was directly compared to that of a GSM trained only on the 200 design training set (without pre-training). The distributions can be seen in Figure \ref{tlVizComp}. Pre-training on related source models reduces the average MAE across the test set by 70\% and the standard deviation by 54\%, resulting in a more accurate and robust surrogate.

\begin{figure*}
\centerline{\includegraphics[width=6.5in]{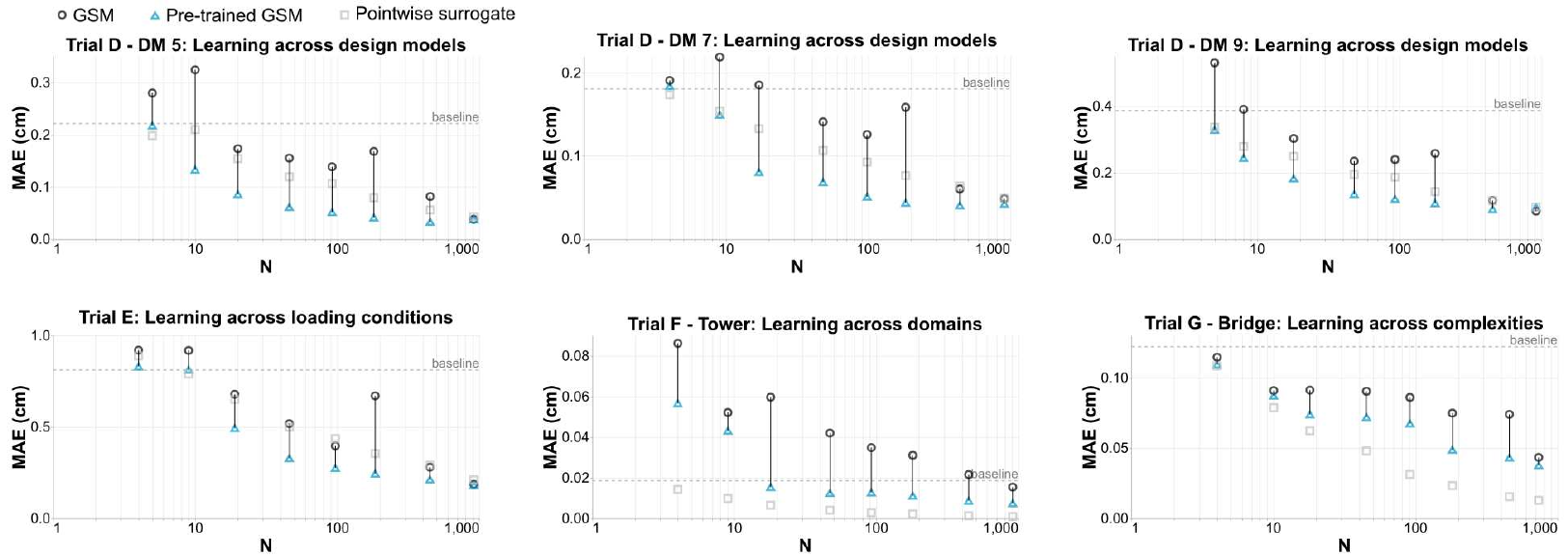}}
\caption{Transfer learning consistently improved the GSMs data efficiency, reducing the amount of training data required to achieve a given prediction accuracy. The baseline refers to a naive model which always predicts the mean displacement from the 1,000 design training set.}
\label{tlAll}
\end{figure*}

\begin{figure} 
\centerline{\includegraphics[width=3.25in]{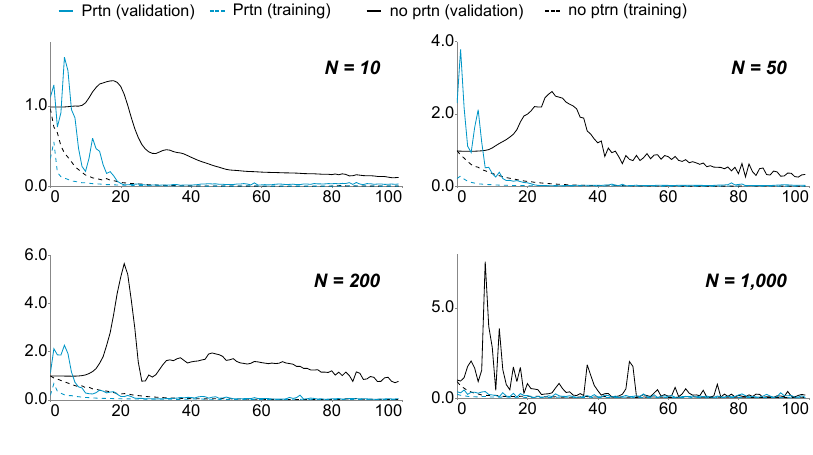}}
\caption{Pre-trained (prtn) GSMs converge faster and to a lower loss value than those trained from scratch, particularly when the training size (N) is small. All curves taken from \emph{\small Trial D}, design model 7 (DM7).}
\label{trainCurves}
\end{figure}

\subsection{Studying data efficiency}
Effective surrogate models should achieve a useful level of prediction accuracy with a minimum amount of training data. The amount of acceptable prediction error and available data are both application dependent. Data efficiency is particularly important in engineering design, where quality design data is often scarce or prohibitively expensive to generate. On the other hand, deep learning methods, with their large number of trainable parameters, are notorious for requiring large data sets. This section explores the effects of transfer learning on the GSM's data efficiency.

\emph{Trial D} was repeated for a variety of target data set sizes. Each data set was generated as in section \ref{dataGen}, and the full 1,000-design set was reserved for testing. To ensure that all data sets were similarly distributed, any designs with maximum displacements exceeding the 90th percentile from the test set were discarded. A different random seed was used in sampling the 1,000-design training set to ensure that training and testing sets did not overlap. In addition to the pre-trained GSM from \emph{Trial D}, a second (not pre-trained) GSM and a pointwise surrogate were trained on the target sets for comparison.

The mean absolute prediction errors as a function of training set size (\emph{N}) can be seen in the top row of Figure \ref{tlAll}. For all models, prediction error correlates negatively with training size, which is expected. In nearly all cases, the pre-trained GSM achieves the lowest prediction errors, followed by the pointwise surrogate and finally by the GSM trained from scratch. Transfer learning improves prediction MAEs by 48.6\%, 40.0\% and 34.1\% for DMs 5, 7, and 9 respectively. The implication is that the amount of training data required to achieve a given predictive performance is reduced by roughly one or two orders of magnitude. For DM 5, a pre-trained GSM requires only 200 designs to achieve an MAE that is within 10\% of the MAE produced by training on 1,000 designs. For DM7, just 100 designs were sufficient to achieve a similar result.

Interestingly, transfer learning was most beneficial for the medium-sized training sets. It is presumed that the smallest training sets do not sufficiently represent the differences between source and target distributions, while the largest training sets are sufficiently large to train a GSM to its predictive limit from scratch. Positive transfer was observed across all design models and training sizes, with the exception of the largest training set for DM9 in which transfer learning increased MAE by 13.7\%. This was the only observed case of negative transfer throughout all trials.

The loss histories of both GSMs reveal further insights about the effects of transfer learning. Figure \ref{trainCurves} shows the evolution of training and validation losses for both GSMs, plotted for four training set sizes. Note that the validation losses for the transfer learned GSM at epoch zero are initially high and comparable to an untrained model. At this point, the conditions are quite similar to those in \emph{Trial C}: the model is attempting to generalize to an unseen topology. However as training progresses, the transfer learned validation losses converge faster and to a lower value than those of the models trained from scratch. Roughly 30 epochs are sufficient to re-train a model, compared to 100 epochs without transfer learning, representing further computational savings.

Encouraged by the positive transfer observed in \emph{Trial D}, one might ask “for which source and target data sets is transfer learning useful?” The design models DM5-9 differ in topology but have the same outer profile, supports and loads. The following trials were designed to test other source/target differences that might occur in a design process. In \emph{Trial E}, a GSM is pre-trained on 1,000 designs from DM 7 and re-trained on identical geometry but with different loads (point-loads at the ends as opposed to a uniform load across the top). Thus \emph{Trial E}, tests the ability to transfer-learn across load cases. The remaining two trials test the ability to transfer-learn across domains. In \emph{Trial F}, a GSM is again pre-trained on DM 7 and retrained on a set of trussed towers. The towers were generated by sampling three handcrafted design parameters. Each is pinned at the bottom and loaded horizontally on the remaining joints. The spanning trusses (DM5-9) and towers differ in topology and outer profile, but have a similar number of bars (27 and 26, respectively). In \emph{Trial G}, the GSM is pre-trained on DM 7 and re-trained on a set of densely trussed bridges. The bridges each consist of 404 bars, making them significantly more complex than the trusses. The bridges are uniformly loaded across the top and simply supported at the bottom. The hyperparameters described in section \ref{training} were used for all trials with the exception of \emph{Trial G}, which used a batch size of 128 and learning rate of 5\(\times10^{-4}\).

The results from \emph{Trials} \emph{E}, \emph{F} and \emph{G} are shown in the bottom row of Figure \ref{tlAll}. In \emph{Trial E}, pre-training on the same geometry but different load cases improved MAE by an average of 25.5\% across all training sizes. In \emph{Trial F}, pre-training on trusses improved MAE predictions on towers by an average of 54.1\%, and in \emph{Trial G}, the same process improved MAE predictions on bridges by 19.8\%. The result is a significant reduction in the amount of required training data. For example, in \emph{Trial E}, a GSM pre-trained on trusses achieves better prediction accuracy when re-trained on 20 tower designs as a GSM which was trained on 500 towers from scratch. Table \ref{resTable} summarizes the findings from \emph{Trials D-G}. Positive transfer was observed across all trials and training sizes, although to varying degrees. As before, the medium-sized training sets generally showed largest benefit and the smallest and largest data sets showed the least. These results further motivate the use of transfer learning to repurpose design data and surrogate models for new tasks.

\begin{table}
\renewcommand{\arraystretch}{1}
\caption{The difference in mean absolute error (\(\Delta\) MAE) between the pre-trained GSM and a GSM trained from scratch, averaged across all training sizes. Transfer learning improved prediction accuracy by 19-48\%.}
\begin{center}
\label{resTable}
\begin{tabular}{c c c l l}
\hline
Trial & Target & \(\Delta\) MAE (cm) & \(\Delta\) MAE \%\\ 
\hline
D & DM5 & -0.087 & -48.6\%\\
D & DM7 & -0.0586 & -40.0\%\\
D & DM9 & -0.106 & -34.1\%\\
E & End Loads & -0.148 & -25.5\%\\
F & Tower & -0.0219 & -54.1\%\\
G & Bridge & -0.0157 & -19.8\%\\
\hline
\end{tabular}
\end{center}
\end{table}

\section{Conclusions and future work} \label{conc}
The proposed Graph-based Surrogate models (GSMs) learn to predict displacement fields given a structure's geometry, supports and loads as inputs. Since the GSM does not rely on handcrafted design parameters, it can be trained on data from multiple design spaces simultaneously, and often benefits from doing so. Transfer learning was presented as an effective method for repurpose GSMs to new tasks by leveraging historical data. GSMs that are pre-trained on a related data set achieve 19-48\% lower prediction errors than those trained from scratch. The result is a more flexible, general and data-efficient surrogate model for trusses.

Future work could consider the increasingly wide array of graph-based learning methods and assess their suitability for trusses. A similar analysis could be performed for surface and volumetric meshes. Though both are easily represented as graphs, meshes differ from trusses in that the topology is not physically meaningful. In terms of transfer learning, further work is required to be able to predict the most effective sources for a given target. One might also explore alternative transfer learning strategies in which learnable parameters are added or frozen during re-training. The ability to learn across designs of varying complexity (\emph{Trial G}) might support hierarchical learning strategies in which models are progressively trained on higher complexity designs. Finally, future work could explore ways of making GSMs generalize to unseen topologies, perhaps by leveraging alternative sources of training data like design competitions for more geometrically diverse data sets.

\begin{acknowledgment}
This research was supported by the Engineering Data Science group at Altair Engineering Inc. and is based upon work supported by the National Science Foundation under Grant No. 1854833. Also a special thanks to Renaud Danhaive and Yijiang Huang for their mentorship.
\end{acknowledgment}

\bibliographystyle{asmems4}
\bibliography{references}

\onecolumn 
\appendix       
\section*{Appendix: GSM tuning results}

\setlength{\intextsep}{7pt}

\begin{table*}[h!]
\renewcommand{\arraystretch}{1}
\caption{Grid search results used to tune the GSMs architecture, learning rate, and number of attention heads. A9 corresponds to \emph{\small L16/C32/C64/C128/C256/C256/C128/L64/L2}, A10 corresponds to \emph{\small L16/C32/C64/C128/C256/C512/C256/C128/L64/L2} and A11 corresponds to \emph{\small L16/C32/C64/C128/C256/C512/C512/C256/C128/L64/L2}. The selected configuration (bold) produced the second-lowest mean squared error (MSE) and had twice as many learnable parameters (and hence more expressivity) as the configuration with the lowest (underlined). The training and testing data were the same as in section \ref{training}. Each configuration was averaged over three trials on a single Tesla K80 GPU.}
\begin{center}
\label{hyperparamTable}
\begin{tabular}{c l c r l l l}
\hline
Architecture & Learning Rate & \# of Heads & Parameters & Train Time (s) & Train MSE (cm\textsuperscript{2}) & Test MSE (cm\textsuperscript{2})\\
\hline
A9 & 1\(\times10^{-4}\) & 4 & 581406 & 1.11\(\times10^{2}\) & 7.44\(\times10^{-3}\) & 1.59\(\times10^{-2}\) \\ 
A9 & 1\(\times10^{-4}\) & 8 & 1151734 & 1.19\(\times10^{2}\) & 6.75\(\times10^{-3}\) & 1.30\(\times10^{-2}\) \\ 
A9 & 1\(\times10^{-4}\) & 12 & 1722062 & 1.30\(\times10^{2}\) & 8.70\(\times10^{-3}\) & 1.52\(\times10^{-2}\) \\ 
A9 & 1\(\times10^{-3}\) & 4 & 581406 & 1.07\(\times10^{2}\) & 5.51\(\times10^{-3}\) & 9.13\(\times10^{-3}\) \\ 
A9 & 1\(\times10^{-3}\) & 8 & 1151734 & 1.15\(\times10^{2}\) & 5.37\(\times10^{-3}\) & 9.25\(\times10^{-3}\) \\ 
A9 & 1\(\times10^{-3}\) & 12 & 1722062 & 1.26\(\times10^{2}\) & 5.88\(\times10^{-3}\) & 9.23\(\times10^{-3}\) \\ 
A9 & 1\(\times10^{-2}\) & 4 & 581406 & 1.04\(\times10^{2}\) & 1.02\(\times10^{-2}\) & 1.26\(\times10^{-2}\) \\ 
A9 & 1\(\times10^{-2}\) & 8 & 1151734 & 1.15\(\times10^{2}\) & 1.03\(\times10^{-2}\) & 1.25\(\times10^{-2}\) \\ 
A9 & 1\(\times10^{-2}\) & 12 & 1722062 & 1.26\(\times10^{2}\) & 1.36\(\times10^{-2}\) & 1.75\(\times10^{-2}\) \\ 
A10 & 1\(\times10^{-4}\) & 4 & 1371426 & 1.31\(\times10^{2}\) & 5.04\(\times10^{-3}\) & 1.32\(\times10^{-2}\) \\ 
A10 & 1\(\times10^{-4}\) & 8 & 2730238 & 1.54\(\times10^{2}\) & 4.83\(\times10^{-3}\) & 1.30\(\times10^{-2}\) \\ 
A10 & 1\(\times10^{-4}\) & 12 & 4089050 & 1.76\(\times10^{2}\) & 4.90\(\times10^{-3}\) & 1.35\(\times10^{-2}\) \\ 
\underline{A10} & \underline{1\(\times10^{-3}\)} & \underline{4} & \underline{1371426} & \underline{1.30\(\times10^{2}\)} & \underline{4.69\(\times10^{-3}\)} & \underline{8.79\(\times10^{-3}\)} \\
\textbf{A10} & \(\mathbf{1\times10^{-3}}\) & \textbf{8} & \textbf{2730238} & \(\mathbf{1.54\times10^{2}}\) & \(\mathbf{5.01\times10^{-3}}\) & \(\mathbf{8.83\times10^{-3}}\) \\ 
A10 & 1\(\times10^{-3}\) & 12 & 4089050 & 1.76\(\times10^{2}\) & 6.45\(\times10^{-3}\) & 1.02\(\times10^{-2}\) \\ 
A10 & 1\(\times10^{-2}\) & 4 & 1371426 & 1.31\(\times10^{2}\) & 9.54\(\times10^{-3}\) & 1.13\(\times10^{-2}\) \\ 
A10 & 1\(\times10^{-2}\) & 8 & 2730238 & 1.53\(\times10^{2}\) & 1.04\(\times10^{-2}\) & 1.28\(\times10^{-2}\) \\ 
A10 & 1\(\times10^{-2}\) & 12 & 4089050 & 1.75\(\times10^{2}\) & 1.56\(\times10^{-2}\) & 1.81\(\times10^{-2}\) \\ 
A11 & 1\(\times10^{-4}\) & 4 & 2423590 & 1.61\(\times10^{2}\) & 5.01\(\times10^{-3}\) & 1.66\(\times10^{-2}\) \\ 
A11 & 1\(\times10^{-4}\) & 8 & 4833030 & 1.99\(\times10^{2}\) & 4.49\(\times10^{-3}\) & 1.42\(\times10^{-2}\) \\ 
A11 & 1\(\times10^{-4}\) & 12 & 7242470 & 2.38\(\times10^{2}\) & 5.13\(\times10^{-3}\) & 1.37\(\times10^{-2}\) \\ 
A11 & 1\(\times10^{-3}\) & 4 & 2423590 & 1.59\(\times10^{2}\) & 5.40\(\times10^{-3}\) & 9.95\(\times10^{-3}\) \\ 
A11 & 1\(\times10^{-3}\) & 8 & 4833030 & 1.99\(\times10^{2}\) & 5.44\(\times10^{-3}\) & 9.51\(\times10^{-3}\) \\ 
A11 & 1\(\times10^{-3}\) & 12 & 7242470 & 2.36\(\times10^{2}\) & 6.72\(\times10^{-3}\) & 1.04\(\times10^{-2}\) \\ 
A11 & 1\(\times10^{-2}\) & 4 & 2423590 & 1.59\(\times10^{2}\) & 1.30\(\times10^{-2}\) & 1.53\(\times10^{-2}\) \\ 
A11 & 1\(\times10^{-2}\) & 8 & 4833030 & 1.97\(\times10^{2}\) & 1.54\(\times10^{-2}\) & 2.00\(\times10^{-2}\) \\ 
A11 & 1\(\times10^{-2}\) & 12 & 7242470 & 2.35\(\times10^{2}\) & 1.99\(\times10^{-2}\) & 2.29\(\times10^{-2}\) \\ 
\hline
\end{tabular}
\end{center}
\end{table*}


\begin{table*}[h!]
\renewcommand{\arraystretch}{1}
\caption{Grid search results used to assess various data transformations applied to the displacements before learning. The training and testing data were the same as in section \ref{training}. Each configuration was averaged over three trials. The standard scaler resulted in the smallest test MSE.} 
\begin{center}
\label{transTable}
\begin{tabular}{c c l l}
\hline
Standard scaler & Log transform & Train MSE (cm\textsuperscript{2}) & Test MSE (cm\textsuperscript{2})\\
\hline
False & False & 5.17\(\times10^{-2}\) & 5.66\(\times10^{-2}\) \\
False & True & 2.48\(\times10^{-2}\) & 2.97\(\times10^{-2}\) \\
\textbf{True} & \textbf{False} & \(\mathbf{3.02\times10^{-3}}\) & \(\mathbf{6.71\times10^{-3}}\) \\
True & True & 7.50\(\times10^{-3}\) & 1.30\(\times10^{-2}\) \\
\hline
\end{tabular}
\end{center}
\end{table*}

\begin{table*}[h!]
\renewcommand{\arraystretch}{1}
\caption{Grid search results used to assess whether it is advantageous to include binary features indicating supports or loads. The training and testing data were the same as in section \ref{training}. Each configuration was averaged over three trials. Note that including support and load information is beneficial despite the fact that all designs are loaded identically.} 
\begin{center}
\label{featureTable}
\begin{tabular}{c c l l}
\hline
Include supports & Include loads & Train MSE (cm\textsuperscript{2}) & Test MSE (cm\textsuperscript{2})\\
\hline
False & False & 4.40\(\times10^{-3}\) & 1.08\(\times10^{-2}\) \\
False & True & 5.49\(\times10^{-3}\) & 1.12\(\times10^{-2}\) \\
True & False & 6.04\(\times10^{-3}\) & 1.02\(\times10^{-2}\) \\
\textbf{True} & \textbf{True} & \(\mathbf{4.65\times10^{-3}}\) & \(\mathbf{9.38\times10^{-3}}\) \\
\hline
\end{tabular}
\end{center}
\end{table*}

\end{document}